\documentclass[11pt,a4paper]{article}
\pdfoutput=1
\usepackage{jheppub}
\usepackage{amsmath,amssymb,bm}
\usepackage{booktabs}
\usepackage{multirow}
\usepackage{xcolor}

\title{Phase Structure and Gravitational-Wave Phenomenology
of a Thermal First-Order Phase Transition}

\author[a]{Gayatri Ghosh}

\affiliation[a]{Department of Physics, Cachar College,
Silchar, Assam, India}

\emailAdd{gayatrighsh@gmail.com}

\abstract{We investigate the phase structure and gravitational-wave (GW)
phenomenology of a cosmological first-order phase transition
described by the finite-temperature effective potential
$V(\phi,T)=D(T^2-T_0^2)\phi^2-ET\phi^3+\lambda\phi^4/4$.
We derive the critical temperature and the broken-phase order
parameter and identify the dimensionless combination
$E^2/(D\lambda)$ that controls the critical-temperature shift.
We then construct a dense numerical atlas containing
$30\,000$ parameter points and map the resulting transition
parameters onto the characteristic GW frequency and peak amplitude.
The scan resolves the multidimensional correlations among
$T_*$, $\alpha$, $\beta/H_*$, $v_w$, $f_{\rm peak}$ and
$\Omega_{\rm GW}^{\rm peak}h^2$.

The present analysis is phenomenological: $T_*$ is defined by the
prescription $T_*=0.95T_c$, while $\beta/H_*$ and $v_w$ are treated
as scan inputs. Consequently, the resulting GW signals are not
interpreted as first-principles predictions. We identify the
additional ingredients required for a predictive calculation,
including the thermal bounce action, nucleation and percolation
temperatures, the transition duration and a microscopic treatment
of bubble-wall friction. The resulting framework provides a systematic numerical characterization of the connection between the phase structure of the finite-temperature potential and the corresponding phenomenological GW parameter space.}
\keywords{Cosmological phase transitions,
Gravitational waves,
Early Universe,
Beyond the Standard Model}

\begin{document}

\maketitle

\section{Introduction}
\label{sec:introduction}

The observation of gravitational waves has opened a new observational
window on fundamental physics. In addition to astrophysical sources,
processes occurring in the early Universe can generate stochastic
gravitational-wave backgrounds. Among the most extensively studied
cosmological sources are first-order phase transitions, in which expanding
bubbles of a lower-energy phase nucleate and subsequently interact with the
surrounding plasma
\cite{Caprini2016,Caprini2020,Athron2024}.

A cosmological first-order phase transition can generate gravitational
radiation through several mechanisms, including bubble-wall collisions,
sound waves in the plasma and magnetohydrodynamic
turbulence. For transitions occurring in a radiation-dominated Universe,
the acoustic contribution is often an important component of the predicted
signal. The detailed properties of the resulting spectrum depend on the
transition temperature, the strength of the transition, the inverse
duration of the transition and the hydrodynamic properties of the plasma
\cite{Hindmarsh2017,RoperPol2024}.

The connection between particle physics and gravitational-wave observables
is nevertheless non-trivial. A microscopic finite-temperature effective
potential must first be used to determine the phase structure and the
tunnelling rate. The resulting nucleation and percolation dynamics then
determine the relevant transition parameters, which must subsequently be
combined with an appropriate description of the plasma and gravitational
wave source. Recent studies have emphasized the importance of these
different stages and the associated theoretical uncertainties
\cite{Athron2024,Caprini2025,Athron2025}.

In this work we investigate a simple finite-temperature scalar potential
of the form
\begin{equation}
V(\phi,T)
=
D(T^2-T_0^2)\phi^2
-
ET\phi^3
+
\frac{\lambda}{4}\phi^4 .
\label{eq:potential}
\end{equation}

This potential provides a transparent phenomenological description of a
thermal first-order phase transition. The temperature-dependent quadratic
term controls the curvature of the symmetric phase, while the cubic term
creates a potential barrier between the symmetric and broken minima.

The central objective of this work is to establish an analytic-to-observable
connection between the microscopic parameters of the finite-temperature
potential and the resulting gravitational-wave phenomenology. In particular,
we identify the dimensionless combination
\begin{equation}
\xi \equiv \frac{E^2}{D\lambda},
\end{equation}
as the principal control parameter governing the shift of the critical
temperature,
\begin{equation}
\frac{T_c}{T_0}=\frac{1}{\sqrt{1-\xi}},
\end{equation}
within the present effective-potential description. This analytic structure
provides a compact characterization of how the thermal barrier modifies the
phase-transition scale. We then propagate this dependence through the
thermodynamic quantities characterizing the transition, in particular
$\Delta V$, $\Delta\rho$ and $\alpha$, and ultimately into the observable
gravitational-wave frequency--amplitude plane.

To resolve the resulting multidimensional structure rather than relying on a
small number of selected benchmark points, we perform a dense numerical scan
of 30,000 parameter points. The scan is therefore used not simply to increase
the sampling density, but to test and quantify the analytic structure
identified above and to determine how the control parameter $\xi$ correlates
with the transition strength and gravitational-wave observables. In this way,
the analysis provides a systematic bridge
\begin{equation}
(D,E,\lambda,T_0)
\longrightarrow
\xi
\longrightarrow
(T_c,\phi_c,\alpha)
\longrightarrow
(f_{\rm peak},\Omega_{\rm GW}^{\rm peak}h^2).
\end{equation}. The resulting scan is used to
quantify the correlations between the microscopic parameters of the
finite-temperature potential and the transition and gravitational-wave
observables, thereby identifying regions of the phenomenological parameter
space associated with stronger gravitational-wave signals. Each point
corresponds to a specific parameter tuple
\begin{equation}
\left(
D,E,\lambda,T_0,\frac{\beta}{H_*},v_w
\right),
\end{equation}
from which the relevant transition and gravitational-wave quantities are
calculated.

The principal transition and gravitational-wave quantities considered in
this work are
\begin{equation}
T_*,
\qquad
\alpha,
\qquad
\frac{\beta}{H_*},
\qquad
v_w,
\qquad
f_{\rm peak},
\qquad
\Omega_{\rm GW}^{\rm peak}h^2 .
\end{equation}

A central feature of the present analysis is the distinction between
quantities calculated directly from the finite-temperature potential and
phenomenological transition parameters. In particular, $\beta/H_*$ and
$v_w$ are treated as scan inputs. They are not claimed to have been
derived from the thermal bounce or from a microscopic plasma-friction
calculation.

We find that the scanned gravitational-wave parameter space exhibits clear
correlations between the transition temperature, transition strength,
inverse transition duration, and wall velocity. In particular, stronger
transitions enhance the peak amplitude, while increasing
$\beta/H_*$ shifts the peak towards higher frequencies and suppresses its
amplitude. We use these correlations to characterize the phenomenological
gravitational-wave parameter space of the model.

This distinction is important because a fully predictive calculation
requires the thermal tunnelling action
\begin{equation}
S_3(T)
=
4\pi
\int_0^\infty dr\,r^2
\left[
\frac{1}{2}
\left(
\frac{d\phi}{dr}
\right)^2
+
V(\phi,T)
-
V(\phi_{\rm false},T)
\right],
\label{eq:S3}
\end{equation}
from which the thermal nucleation rate and, together with the nucleation
probability, the nucleation and percolation temperatures can be determined;
the inverse duration can then be obtained from the temperature dependence
of $S_3(T)/T$.
First-order cosmological phase transitions provide a well-established mechanism
for generating stochastic gravitational-wave backgrounds and therefore offer a
possible connection between early-Universe particle physics and gravitational-wave
observations \cite{Caprini:2016yrh,Caprini:2019egz,Athron:2023xlr}.
The resulting gravitational-wave signal can receive contributions from bubble-wall
collisions, sound waves in the primordial plasma, and magnetohydrodynamic
turbulence, with the acoustic contribution playing an important role for a broad
class of transitions \cite{Hindmarsh:2017gnf,Caprini:2019egz}.
The spectral amplitude and shape depend on the transition temperature, transition
strength, inverse duration, and hydrodynamic properties of the plasma
\cite{Caprini:2016yrh,Caprini:2019egz,Athron:2023xlr}.
Recent developments have also emphasized that the finite lifetime and nonlinear
evolution of the acoustic source can modify the gravitational-wave amplitude,
particularly for stronger transitions \cite{RoperPol:2023xwe,Caprini:2024xx}.
A fully predictive treatment further requires a dynamical calculation of the
finite-temperature effective potential, tunnelling rate, nucleation and
percolation temperatures, and transition duration. Numerical tools such as
PhaseTracer2 provide a framework for connecting finite-temperature effective
potentials with phase-transition and gravitational-wave observables
\cite{Athron:2024dbx}. Gravitational waves provide a complementary probe of early-Universe
physics, with stochastic backgrounds arising from a variety of
cosmological sources. Recent studies have explored such connections in
different particle-physics settings, including domain-wall dynamics
\cite{Ghosh:2026ddg}.
 The organization of the paper is as follows. In section
\ref{sec:model} we discuss the finite-temperature model and derive its
phase structure. Section \ref{sec:thermo} introduces the thermodynamic
parameters characterizing the transition. Section \ref{sec:gw} describes
the gravitational-wave parameterization used in the numerical analysis.
Section \ref{sec:scan} presents the parameter scan and benchmark dataset.
Section \ref{sec:results} discusses the resulting high-density
parameter-space correlations. Section \ref{sec:limitations} discusses the
limitations of the present phenomenological treatment and the required
extensions toward a first-principles calculation. We conclude in section
\ref{sec:conclusion}.

\section{Finite-temperature model}
\label{sec:model}

We consider a real scalar order parameter $\phi$ described by the
finite-temperature potential
\begin{equation}
V(\phi,T)
=
D(T^2-T_0^2)\phi^2
-
ET\phi^3
+
\frac{\lambda}{4}\phi^4 .
\end{equation}

The parameters are assumed to satisfy
\begin{equation}
D>0,
\qquad
E>0,
\qquad
\lambda>0.
\end{equation}
The stationary points of the potential satisfy
\begin{equation}
\frac{\partial V}{\partial\phi}
=
2D(T^2-T_0^2)\phi
-
3ET\phi^2
+
\lambda\phi^3
=0 .
\label{eq:stationary}
\end{equation}

The symmetric solution is
\begin{equation}
\phi=0.
\end{equation}

The non-zero extrema are obtained from
\begin{equation}
\lambda\phi^2
-
3ET\phi
+
2D(T^2-T_0^2)
=0,
\end{equation}
giving
\begin{equation}
\phi_{\pm}(T)
=
\frac{
3ET
\pm
\sqrt{
9E^2T^2
-
8D\lambda(T^2-T_0^2)
}
}
{2\lambda}.
\label{eq:phipm}
\end{equation}

The nature of the stationary points is determined by the second derivative,
\begin{equation}
\frac{\partial^2 V}{\partial\phi^2}
=
2D(T^2-T_0^2)-6ET\phi+3\lambda\phi^2.
\end{equation}
\subsection{Critical temperature}

The critical temperature is defined by the degeneracy condition
\begin{equation}
V(0,T_c)
=
V(\phi_c,T_c),
\label{eq:degeneracy}
\end{equation}
where $\phi_c$ denotes the non-zero minimum at the critical temperature.

At the critical temperature, the broken minimum satisfies both the
stationarity condition and the degeneracy condition. The stationarity
condition gives
\begin{equation}
2D(T_c^2-T_0^2)
-3ET_c\phi_c
+\lambda\phi_c^2
=0,
\label{eq:stationary_tc}
\end{equation}
while the degeneracy condition,
$V(0,T_c)=V(\phi_c,T_c)$, gives
\begin{equation}
D(T_c^2-T_0^2)
-ET_c\phi_c
+\frac{\lambda}{4}\phi_c^2
=0.
\label{eq:degeneracy_tc}
\end{equation}

Multiplying Eq.~\eqref{eq:degeneracy_tc} by $2$ and subtracting it
from Eq.~\eqref{eq:stationary_tc} yields
\begin{equation}
-ET_c\phi_c
+\frac{\lambda}{2}\phi_c^2
=0.
\end{equation}
For the non-zero broken minimum, this gives
\begin{equation}
\phi_c
=
\frac{2ET_c}{\lambda}.
\label{eq:phic_derivation}
\end{equation}

Substituting this result into the degeneracy condition gives
\begin{equation}
D(T_c^2-T_0^2)
-\frac{2E^2T_c^2}{\lambda}
+\frac{E^2T_c^2}{\lambda}
=0,
\end{equation}
and hence
\begin{equation}
T_c^2
\left(
1-\frac{E^2}{D\lambda}
\right)
=
T_0^2.
\end{equation}
Therefore, the critical temperature is
\begin{equation}
T_c
=
\frac{T_0}
{\sqrt{1-\dfrac{E^2}{D\lambda}}}.
\label{eq:Tc}
\end{equation}

It is useful to introduce the dimensionless combination
\begin{equation}
\xi
\equiv
\frac{E^2}{D\lambda}.
\label{eq:xi}
\end{equation}
The critical temperature can then be written as
\begin{equation}
\frac{T_c}{T_0}
=
\frac{1}{\sqrt{1-\xi}},
\label{eq:Tc_over_T0}
\end{equation}
or equivalently,
\begin{equation}
\boxed{
\xi
=
1-\left(\frac{T_0}{T_c}\right)^2
}.
\label{eq:xi_tc}
\end{equation}

The corresponding broken-phase value is
\begin{equation}
\phi_c
=
\frac{2ET_c}{\lambda},
\label{eq:phic}
\end{equation}
and hence
\begin{equation}
\frac{\phi_c}{T_c}
=
\frac{2E}{\lambda}.
\label{eq:phicratio}
\end{equation}

The parameter region used in the numerical scan satisfies
\begin{equation}
0
<
\frac{E^2}{D\lambda}
<
1.
\label{eq:validdomain}
\end{equation}
For the present parameterization, the condition
$E^2/(D\lambda)<1$ ensures that the critical temperature is finite and
real. The scan is restricted to $E>0$, so that the cubic term provides
the barrier characteristic of the first-order transition.
\subsection{The analytic control parameter $\xi$}

The critical temperature admits a particularly transparent interpretation in
terms of the dimensionless combination
\begin{equation}
\xi \equiv \frac{E^2}{D\lambda}.
\end{equation}
The critical-temperature relation can then be written as
\begin{equation}
T_c = \frac{T_0}{\sqrt{1-\xi}},
\end{equation}
or equivalently,
\begin{equation}
\frac{T_c}{T_0}=(1-\xi)^{-1/2}.
\end{equation}
Thus, $\xi$ measures the relative importance of the thermal cubic term
responsible for the barrier compared with the quadratic and quartic
coefficients of the effective potential.

For $\xi\ll1$, the critical-temperature shift is approximately
\begin{equation}
\frac{T_c}{T_0}
\simeq 1+\frac{\xi}{2}
+\mathcal{O}(\xi^2),
\end{equation}
whereas the sensitivity to $\xi$ increases as $\xi$ approaches the boundary
of the parameter region for which the broken and symmetric minima admit the
critical-temperature solution. The same parameter therefore provides a
convenient one-dimensional characterization of part of the otherwise
multidimensional parameter dependence.

\begin{figure}[htp]
    \centering
    \includegraphics[width=\textwidth]{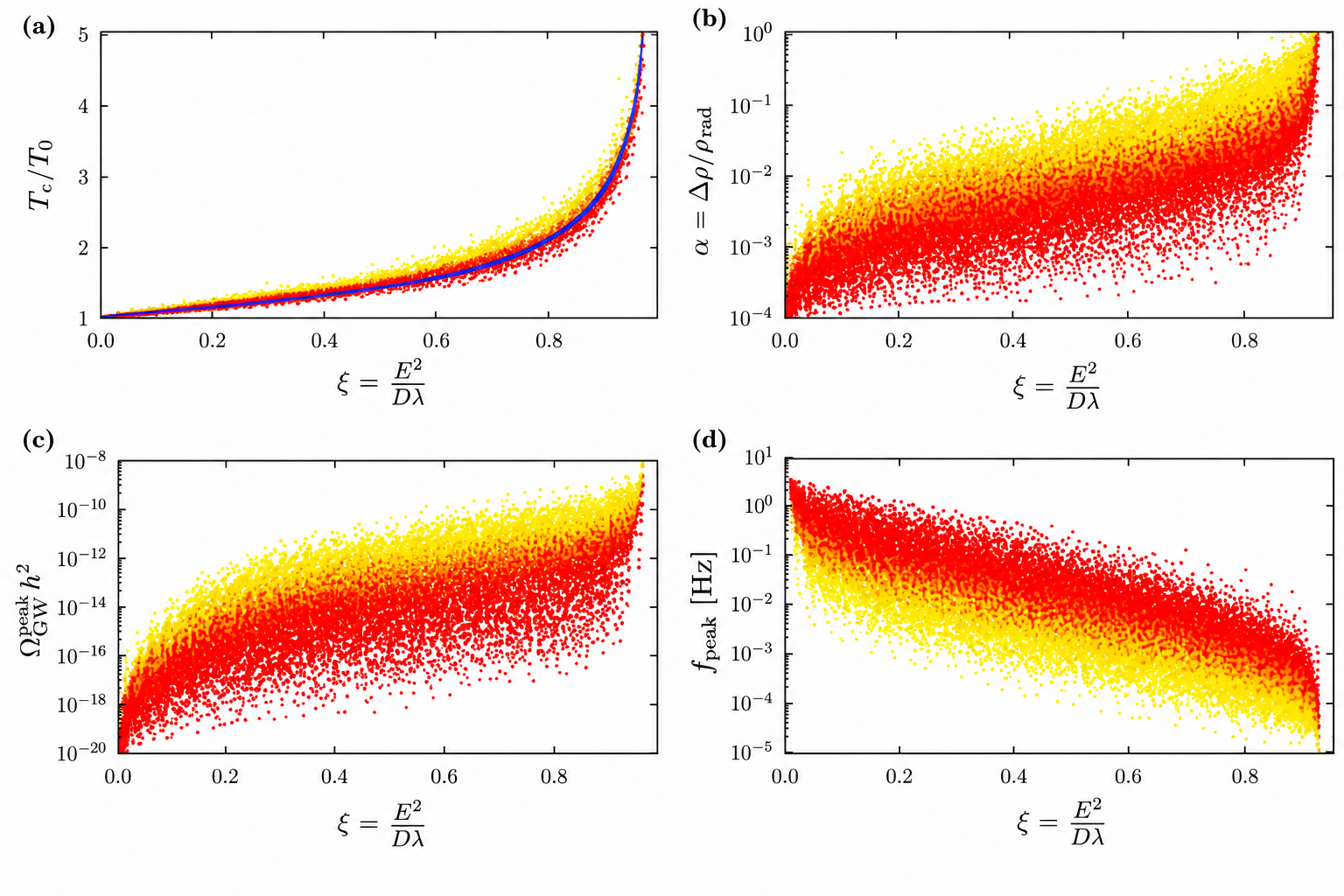}
    \caption{Analytic and numerical correlations controlled by the
    dimensionless parameter $\xi=E^2/(D\lambda)$. The four panels show
    (a) $T_c/T_0$ as a function of $\xi$, together with the analytic
    relation $T_c/T_0=(1-\xi)^{-1/2}$, (b) $\alpha$ as a function of $\xi$,
    colour-coded by $E/\lambda$, (c) $\Omega_{\rm GW}^{\rm peak}h^2$ as a
    function of $\xi$, colour-coded by $\beta/H_*$, and (d) $f_{\rm peak}$
    as a function of $\xi$.}
    \label{fig:xi_control}
\end{figure}

It is important, however, that $\xi$ does not by itself completely determine
the transition strength. The order parameter at the critical temperature is
controlled by
\begin{equation}
\frac{\phi_c}{T_c}=\frac{2E}{\lambda},
\end{equation}
while the released energy density and the gravitational-wave amplitude depend
on the full temperature-dependent potential. The role of $\xi$ in the present
work is consequently to provide an analytic organizing variable whose
correlations with the thermodynamic and gravitational-wave observables are
tested explicitly in the numerical scan.
The dimensionless combination
\begin{equation}
\xi=\frac{E^2}{D\lambda}
\end{equation}
provides the central analytic control parameter of the phase-transition
dynamics considered in this work. In terms of $\xi$, the critical temperature
is determined by
\begin{equation}
\frac{T_c}{T_0}=\frac{1}{\sqrt{1-\xi}}.
\end{equation}
Figure~\ref{fig:xi_control} establishes how this analytic dependence is
reflected in the numerical parameter space and propagated into the
thermodynamic and gravitational-wave observables. The $T_c/T_0$--$\xi$
relation directly verifies the analytic prediction, while the correlations
of $\alpha$, $\Omega_{\rm GW}^{\rm peak}h^2$ and $f_{\rm peak}$ with $\xi$
reveal how the critical-temperature shift is connected to the observable
properties of the transition. The residual spread in these quantities
reflects their dependence on the remaining independent parameters of the
finite-temperature potential and on the phenomenological transition
parameters.
Figure~\ref{fig:xi_control} establishes the corresponding analytic--numerical
connection across the full parameter scan. The $T_c/T_0$--$\xi$ correlation
reproduces the analytic dependence, while the $\alpha$--$\xi$,
$\Omega_{\rm GW}^{\rm peak}h^2$--$\xi$, and $f_{\rm peak}$--$\xi$
correlations demonstrate how the $\xi$ dependence is propagated through the
thermodynamic properties of the phase transition into the gravitational-wave
observables. The observed spread in these quantities further quantifies the
dependence on the remaining independent parameters of the finite-temperature
potential and on the phenomenological transition parameters.
\subsection{Transition temperature}

For the present phenomenological scan we define the characteristic
transition temperature according to
\begin{equation}
T_*
=
0.95T_c .
\label{eq:Tstar}
\end{equation}

This prescription is used consistently to construct the numerical
dataset. It should not be confused with a bounce-derived nucleation or
percolation temperature.

At $T=T_*$, the broken-phase minimum corresponds to the
$\phi_+$ branch of Eq.~\eqref{eq:phipm}, giving
\begin{equation}
\phi_*
=
\frac{
3ET_*
+
\sqrt{
9E^2T_*^2
-
8D\lambda(T_*^2-T_0^2)
}
}
{2\lambda}.
\label{eq:phistar}
\end{equation}

The scan is restricted to parameter points for which the discriminant
in Eq.~\eqref{eq:phistar} is non-negative, ensuring a real broken-phase
stationary point at $T_*$.
\section{Thermodynamic properties}
\label{sec:thermo}

The free-energy difference between the false and broken phases is
\begin{equation}
\Delta V(T)
=
V(\phi_{\rm false},T)
-
V(\phi_{\rm broken},T).
\label{eq:deltav}
\end{equation}

With the convention adopted in eq.~\eqref{eq:potential},
\begin{equation}
V(0,T)=0,
\end{equation}
and therefore
\begin{equation}
\Delta V(T)
=
-
V(\phi_{\rm broken},T).
\end{equation}

The corresponding energy-density difference is
\begin{equation}
\Delta\rho(T)
=
\Delta V(T)
-
T\frac{d\Delta V(T)}{dT}.
\label{eq:deltarho}
\end{equation}

Equivalently, at the transition temperature,
\begin{equation}
\Delta\rho_*
=
-
V(\phi_*,T_*)
+
T_*
\left.
\frac{\partial V}{\partial T}
\right|_{\phi=\phi_*,T=T_*}.
\label{eq:deltarho2}
\end{equation}

The explicit temperature derivative is
\begin{equation}
\frac{\partial V}{\partial T}
=
2DT\phi^2-E\phi^3.
\label{eq:dVdT}
\end{equation}
Using the explicit form of the potential and evaluating at the broken
minimum, the energy-density difference can be written as
\begin{equation}
\Delta\rho_*
=
-D(T_*^2-T_0^2)\phi_*^2
+E T_*\phi_*^3
-\frac{\lambda}{4}\phi_*^4
+2DT_*^2\phi_*^2
-E T_*\phi_*^3,
\end{equation}
or equivalently,
\begin{equation}
\Delta\rho_*
=
D(T_*^2+T_0^2)\phi_*^2
-\frac{\lambda}{4}\phi_*^4.
\label{eq:deltarho_explicit}
\end{equation}
The transition-strength parameter is defined as
\begin{equation}
\alpha
=
\frac{\Delta\rho_*}
{\rho_{\rm rad}(T_*)},
\label{eq:alpha}
\end{equation}
where
\begin{equation}
\rho_{\rm rad}(T_*)
=
\frac{\pi^2}{30}g_*T_*^4.
\label{eq:rhorad}
\end{equation}

For the present analysis we take
\begin{equation}
g_*=106.75.
\label{eq:gstar}
\end{equation}
The resulting $\alpha$ values are therefore calculated directly from the
finite-temperature scalar potential at the prescribed temperature $T_*$.

\section{Gravitational-wave signal}
\label{sec:gw}
A first-order phase transition can generate gravitational waves through
bubble dynamics and the resulting fluid motion. For the phenomenological
scan, we focus on the gravitational-wave contribution from sound waves
in the plasma
\cite{Caprini2020,Hindmarsh2017,RoperPol2024}.

For the phenomenological scan we use the characteristic acoustic-source
peak frequency
\begin{equation}
f_{\rm peak}
=
1.9\times10^{-5}\,{\rm Hz}
\left(
\frac{\beta/H_*}{v_w}
\right)
\left(
\frac{T_*}{100\,{\rm GeV}}
\right)
\left(
\frac{g_*}{100}
\right)^{1/6}.
\label{eq:fpeak}
\end{equation}

The efficiency factor is parameterized as
\begin{equation}
\kappa_{\rm sw}
=
\frac{\alpha}
{
0.73+0.083\sqrt{\alpha}+\alpha
}.
\label{eq:kappa}
\end{equation}

The peak value of the gravitational-wave energy-density spectrum is then
approximated by
\begin{equation}
\Omega_{\rm GW}^{\rm peak}h^2
=
2.65\times10^{-6}
\left(
\frac{H_*}{\beta}
\right)
\left[
\frac{
\kappa_{\rm sw}\alpha
}{
1+\alpha
}
\right]^2
\left(
\frac{100}{g_*}
\right)^{1/3}
v_w.
\label{eq:omegapeak}
\end{equation}

The above equations are used to map each phase-transition parameter
tuple onto the gravitational-wave frequency--amplitude plane.

It is important to emphasize that eq.~\eqref{eq:omegapeak} is a
phenomenological acoustic-source estimate. Recent studies have shown that
the finite lifetime and nonlinear evolution of the acoustic source can
modify the amplitude, particularly for stronger transitions
\cite{RoperPol2024,Caprini2025}. We therefore interpret the present
results as a parameter-space scan rather than as a precision prediction
for the stochastic background.

\section{Numerical parameter scan and analytic--numerical correlations}
\label{sec:scan}
The numerical scan is designed to test the analytic structure identified in
Section~2 rather than serving solely as a broad phenomenological sampling.
For each parameter point, we evaluate the thermodynamic quantities at the
phenomenological transition temperature and map them onto the corresponding
gravitational-wave observables. In addition to the original potential
parameters $(D,E,\lambda,T_0)$, we therefore retain the derived quantity
\begin{equation}
\xi=\frac{E^2}{D\lambda}
\end{equation}
as an explicit scan variable.

The resulting 30,000-point dataset allows us to investigate whether the
analytic dependence of $T_c$ on $\xi$ is reflected in the numerical
thermodynamic evolution and how this dependence propagates into
$\alpha$, $f_{\rm peak}$ and $\Omega_{\rm GW}^{\rm peak}h^2$.


We perform a high-density numerical scan over the parameter set
\begin{equation}
\left\{
D,E,\lambda,T_0,
\frac{\beta}{H_*},
v_w
\right\}.
\end{equation}

The parameter region is restricted by
\begin{equation}
D>0,
\qquad
\lambda>0,
\qquad
0<
\frac{E^2}{\lambda D}
<1.
\end{equation}

In accordance with the phenomenological setup described above,
$\beta/H_*$ and $v_w$ are treated as independent scan inputs rather than
being derived from the thermal bounce action or a microscopic
plasma-friction calculation.

The benchmark ranges defining the scan are
\begin{align}
0.22 &\leq D \leq 0.46,
\\
0.050 &\leq E \leq 0.205,
\\
0.090 &\leq \lambda \leq 0.340,
\\
95~{\rm GeV}
&\leq T_0
\leq1000~{\rm GeV},
\\
35
&\leq
\frac{\beta}{H_*}
\leq600,
\\
0.50
&\leq v_w
\leq0.82.
\label{eq:ranges}
\end{align}
The parameters are sampled within the ranges specified in
Eqs.~(5.3)--(5.8), subject to the validity condition
$0<E^2/(\lambda D)<1$.
Parameter points that do not satisfy the above validity conditions or
do not yield a real broken-phase stationary point at $T_*$ are discarded.
The remaining sample contains
\begin{equation}
N_{\rm scan}=30\,000
\end{equation}
valid parameter points.

For each point, the quantities $T_c$, $T_*$, $\phi_*$, $\alpha$,
$f_{\rm peak}$ and $\Omega_{\rm GW}^{\rm peak}h^2$ are calculated from
the six input parameters.

The numerical dataset therefore has the structure
\begin{equation}
\mathcal{D}
=
\left\{
D,E,\lambda,T_0,\beta/H_*,v_w,
T_c,T_*,\phi_*,\alpha,
f_{\rm peak},
\Omega_{\rm GW}^{\rm peak}h^2
\right\}_{i=1}^{30000}.
\end{equation}

No statistical interpretation is assigned to the density of points.
The dense sampling is used solely to resolve correlations and
structures within the chosen phenomenological parameter region.
\section{Benchmark parameter points}
\label{sec:benchmarks}
Table~\ref{tab:benchmarks} summarizes representative benchmark points
selected from the numerical scan.
\begin{table}[htp]
\centering
\caption{Representative benchmark points from the numerical dataset. The quantities
$T_c$, $T_\ast$ and $\alpha$ are calculated from the finite-temperature potential,
with $T_\ast=0.95T_c$. The quantity $\beta/H_\ast$ is a phenomenological scan input.}
\label{tab:benchmarks}
\begin{tabular}{lcccccccc}
\toprule
Benchmark & $D$ & $E$ & $\lambda$ & $T_0$ [GeV] &
$T_c$ [GeV] & $T_\ast$ [GeV] & $\alpha$ & $\beta/H_\ast$ \\
\midrule
BP01 & 0.220 & 0.050 & 0.090 & 95   & 101.633 & 96.551  & 0.02642 & 35  \\
BP02 & 0.240 & 0.060 & 0.105 & 120  & 129.615 & 123.134 & 0.02823 & 50  \\
BP03 & 0.260 & 0.072 & 0.120 & 150  & 164.266 & 156.053 & 0.03065 & 70  \\
BP04 & 0.280 & 0.085 & 0.140 & 180  & 199.301 & 189.336 & 0.03160 & 90  \\
BP05 & 0.300 & 0.100 & 0.165 & 230  & 257.473 & 244.599 & 0.03175 & 120 \\
BP06 & 0.320 & 0.115 & 0.185 & 300  & 340.425 & 323.404 & 0.03327 & 160 \\
BP07 & 0.350 & 0.135 & 0.215 & 400  & 459.495 & 436.520 & 0.03508 & 220 \\
BP08 & 0.380 & 0.155 & 0.245 & 550  & 638.524 & 606.598 & 0.03693 & 300 \\
BP09 & 0.420 & 0.180 & 0.290 & 750  & 875.419 & 831.648 & 0.03842 & 420 \\
BP10 & 0.460 & 0.205 & 0.340 & 1000 & 1169.372 & 1110.904 & 0.03940 & 600 \\
\bottomrule
\end{tabular}
\end{table}
\section{Numerical results}
\label{sec:results}

The main result of the numerical analysis is the high-density scan of the
phase-transition and gravitational-wave parameter space.

\begin{figure}[t]
    \centering
    \includegraphics[width=0.99\textwidth]{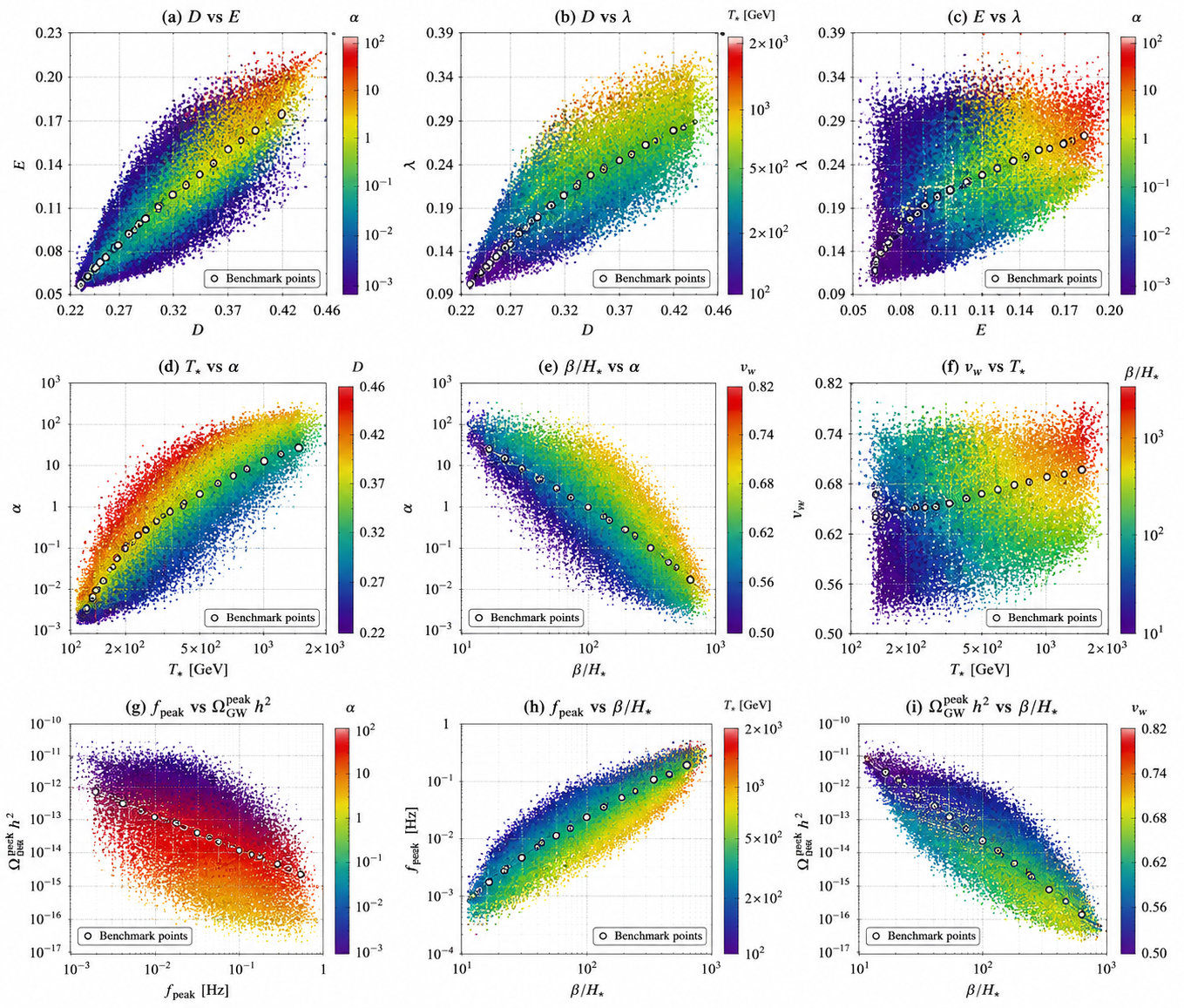}
    \caption{\label{fig:multicolour_scan}
    High-density multicolour numerical scan of the phase-transition and
    gravitational-wave parameter space. The figure shows projections of the
    numerical parameter space in the $(D,E)$, $(D,\lambda)$, $(E,\lambda)$,
    $(T_\ast,\alpha)$, $(\beta/H_\ast,\alpha)$, $(T_\ast,v_w)$,
    $(f_{\rm peak},\Omega_{\rm GW}^{\rm peak}h^2)$,
    $(\beta/H_\ast,f_{\rm peak})$, and
    $(\beta/H_\ast,\Omega_{\rm GW}^{\rm peak}h^2)$ planes.
    The colour scale in each panel represents the third quantity indicated
    by the corresponding colour bar. The open circles denote the
    representative benchmark points. The dense sampling is used to resolve
    correlations within the chosen phenomenological parameter region and
    is not assigned a statistical interpretation.}
\end{figure}

Figure~\ref{fig:multicolour_scan} provides a direct visualization of the
correlations generated by the numerical scan. Panels (a)--(c) show the
correlations among the underlying parameters of the finite-temperature
potential, while panels (d)--(f) display correlations involving the
transition temperature, transition strength, inverse transition duration
and wall velocity. Panels (g)--(i) show the corresponding structure in
the gravitational-wave peak frequency and peak amplitude.

In particular, panel~(h) exhibits the expected increase of the
characteristic frequency with $\beta/H_\ast$, consistent with
\begin{equation}
f_{\rm peak}\propto
\frac{\beta}{H_\ast}\frac{T_\ast}{v_w}.
\end{equation}

Panel~(i) shows the dependence of the peak amplitude on the inverse
transition duration. Within the adopted phenomenological parameterization,
the amplitude scales as
\begin{equation}
\Omega_{\rm GW}^{\rm peak}h^2
\propto
\left(\frac{\beta}{H_\ast}\right)^{-1}
\left(\frac{\kappa_{\rm sw}\alpha}{1+\alpha}\right)^2v_w.
\end{equation}

Thus, for fixed $\alpha$ and $v_w$, increasing $\beta/H_\ast$ suppresses
the peak amplitude, while simultaneously shifting the peak towards higher
frequencies.
The frequency--amplitude plane is governed by the combined dependence
\begin{equation}
f_{\rm peak}
\propto
\frac{\beta}{H_*}
\frac{T_*}{v_w},
\end{equation}
while the peak amplitude scales approximately as
\begin{equation}
\Omega_{\rm GW}^{\rm peak}h^2
\propto
\frac{1}{\beta/H_*}
\left[
\frac{\kappa_{\rm sw}\alpha}
{1+\alpha}
\right]^2
v_w.
\end{equation}

The resulting multidimensional structure therefore reflects the interplay
between the transition strength, inverse transition duration, transition
temperature and wall velocity.

\subsection{Transition strength and gravitational-wave amplitude}

The transition strength parameter $\alpha$ provides a measure of the
released energy relative to the radiation density,
\begin{equation}
\alpha
=
\frac{\Delta\rho_*}{\rho_{\rm rad}(T_*)}.
\end{equation}

An increase in $\alpha$ generally enhances the efficiency of conversion
into bulk fluid motion. This is reflected in the nonlinear dependence
\begin{equation}
\Omega_{\rm GW}^{\rm peak}h^2
\propto
\left[
\frac{\kappa_{\rm sw}\alpha}
{1+\alpha}
\right]^2.
\end{equation}

The numerical scan therefore provides a visualization of how the
transition thermodynamics populate the gravitational-wave
frequency--amplitude plane. The high-density scan shown in
Fig.~\ref{fig:multicolour_scan}, particularly panel~(g), shows the
dependence of the peak gravitational-wave amplitude on the
transition-strength parameter $\alpha$. The colour coding additionally
illustrates the dependence on the phenomenological inverse transition
duration $\beta/H_*$. The resulting distribution reflects the combined
dependence of the gravitational-wave observables on the transition
temperature, transition strength, inverse transition duration, and wall
velocity.

\begin{figure}[htp]
    \centering
    \includegraphics[width=0.98\textwidth]{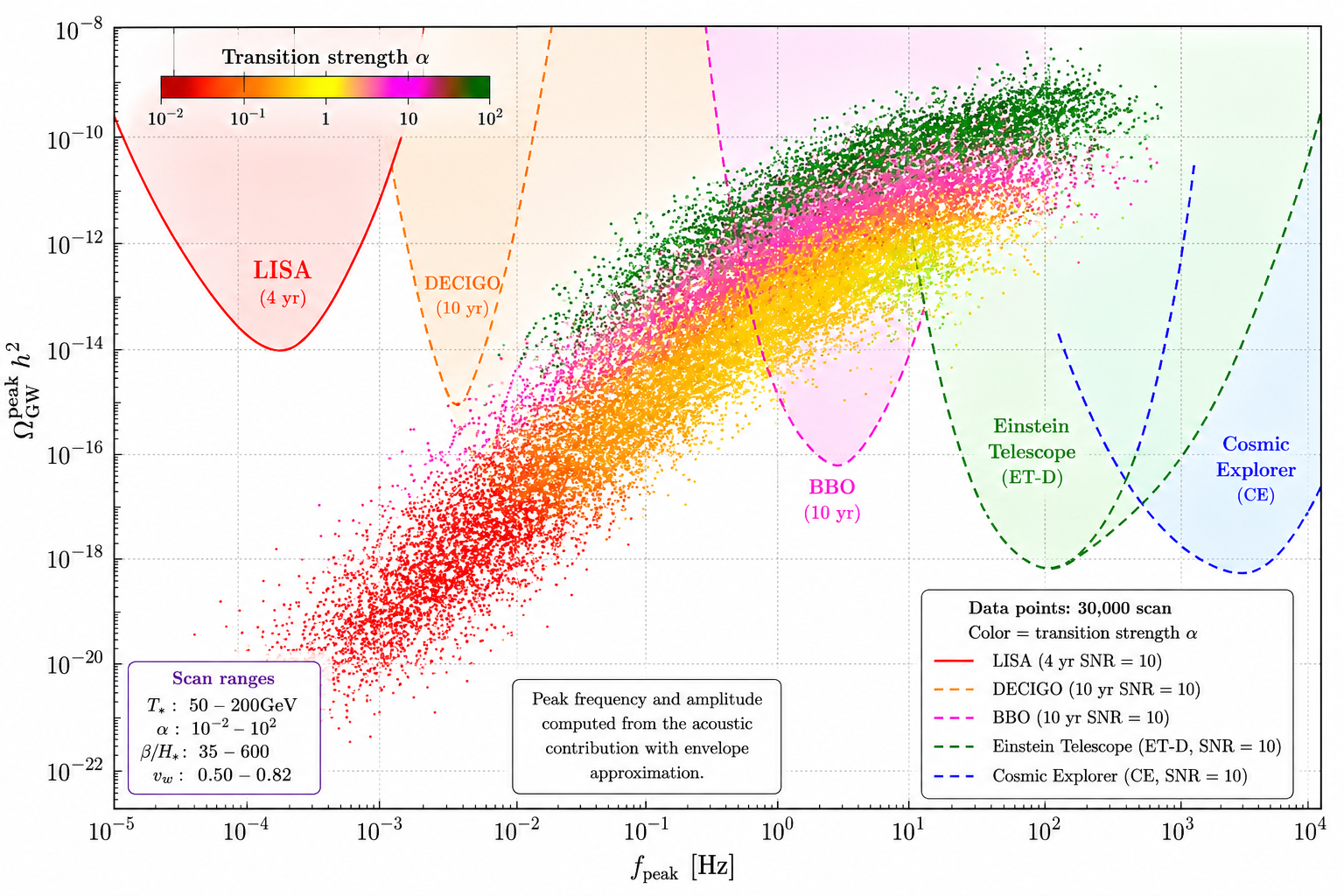}
    \caption{Detector-oriented gravitational-wave parameter space obtained from the 30,000-point phenomenological scan. The peak gravitational-wave frequency $f_{\rm peak}$ is plotted against the peak amplitude $\Omega_{\rm GW}^{\rm peak}h^2$ on logarithmic axes. The colour scale represents the transition-strength parameter $\alpha$. The distribution demonstrates how the thermodynamic transition strength, itself correlated with the analytic control parameter $\xi=E^2/(D\lambda)$, propagates into the observable frequency--amplitude plane. The curves indicate representative
sensitivity regions associated with LISA, DECIGO, BBO, Einstein Telescope (ET-D), and Cosmic Explorer (CE).}
    \label{fig:gw_detector_plane}
\end{figure}

The detector-oriented gravitational-wave parameter space is shown in
Fig.~\ref{fig:gw_detector_plane}. The figure displays the peak frequency
$f_{\rm peak}$ against the peak gravitational-wave amplitude
$\Omega_{\rm GW}^{\rm peak}h^2$ for the complete 30,000-point numerical
scan. The colour coding represents the transition-strength parameter
$\alpha$, thereby showing how the strength of the phase transition
populates the frequency--amplitude plane. The characteristic frequency
follows the approximate scaling
\begin{equation}
f_{\rm peak}
\propto
\frac{\beta}{H_*}
\frac{T_*}{v_w},
\end{equation}
while the peak amplitude is suppressed for increasing $\beta/H_*$ and
enhanced by increasing transition strength, subject to the simultaneous
variation of the other scan parameters.

\begin{figure}[htp]
    \centering
    \includegraphics[width=\textwidth]{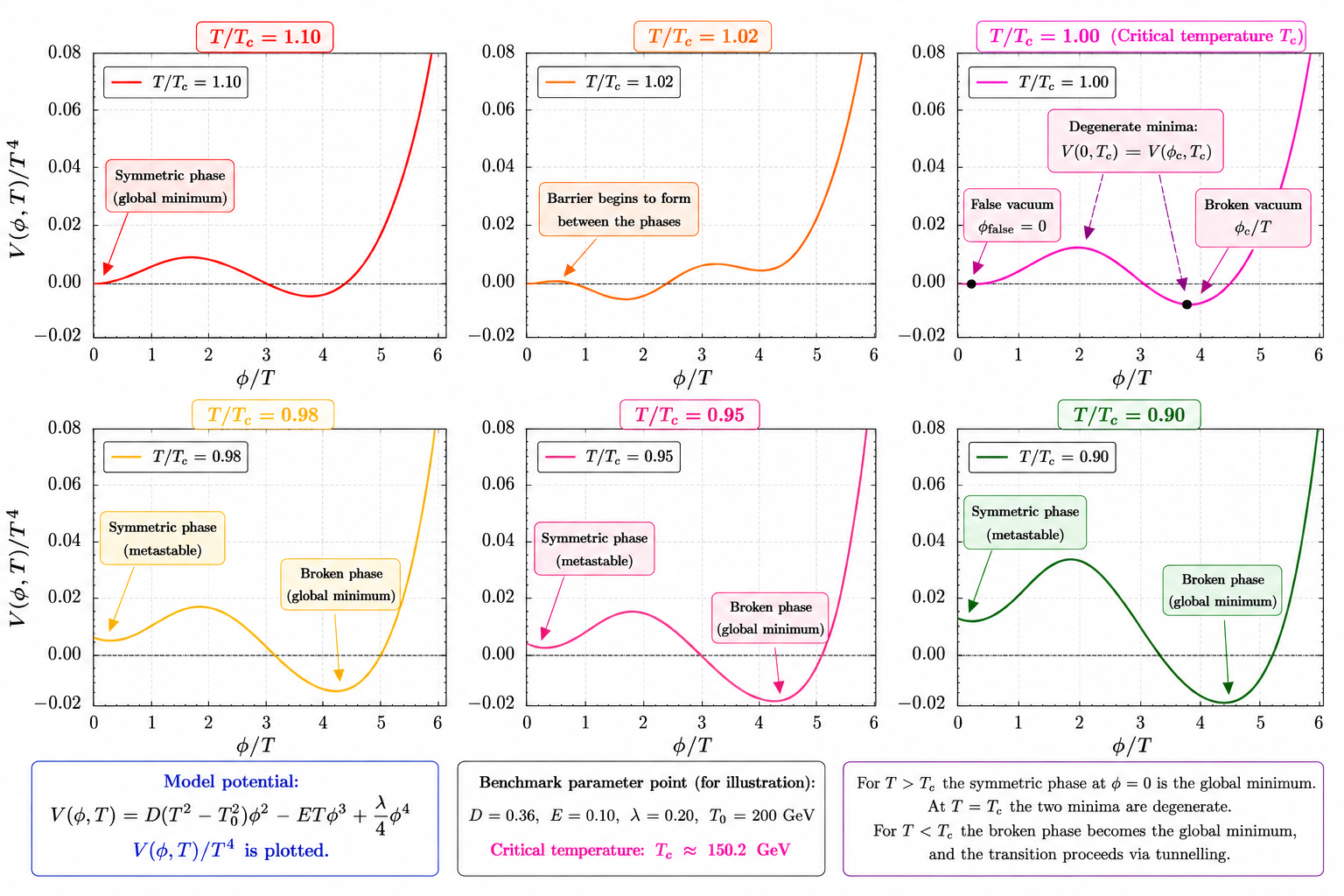}
    \caption{Evolution of the finite-temperature effective potential
    around the critical temperature. The dimensionless potential
    $V(\phi,T)/T^4$ is shown as a function of $\phi/T$ for
    $T/T_c=1.10$, $1.02$, $1.00$, $0.98$, $0.95$, and $0.90$.
    Above the critical temperature the symmetric phase at $\phi=0$
    is energetically preferred. At $T=T_c$ the symmetric and broken
    minima are degenerate, while below $T_c$ the broken phase becomes
    the global minimum and the symmetric phase remains metastable as
    long as the barrier persists. The evolution illustrates the
    barrier structure responsible for the first-order thermal phase
    transition in the phenomenological effective potential
    $V(\phi,T)
    =D(T^2-T_0^2)\phi^2
    -ET\phi^3
    +\frac{\lambda}{4}\phi^4$. The plotted temperature evolution is illustrative of the phase
    structure of the model.}
    \label{fig:effective_potential}
\end{figure}

The thermal origin of the first-order transition can be visualized
directly from the finite-temperature effective potential. Figure~\ref{fig:effective_potential}
shows the evolution of the dimensionless potential
$V(\phi,T)/T^4$ as a function of $\phi/T$ at several temperatures
around the critical temperature. For temperatures above $T_c$, the
symmetric configuration $\phi=0$ is energetically preferred. As the
temperature approaches $T_c$, a barrier separates the symmetric and
broken-phase configurations. At the critical temperature, the two
phases become degenerate, satisfying
\begin{equation}
V(0,T_c)=V(\phi_c,T_c).
\end{equation}
For temperatures below $T_c$, the broken-phase minimum becomes the
global minimum, while the symmetric phase remains metastable as long
as the barrier persists. This temperature evolution provides a direct
illustration of the first-order nature of the transition generated by
the cubic thermal term in the effective potential.

\subsection{Transition duration and peak frequency}

The inverse duration parameter is related to the temperature dependence of
the thermal tunnelling action through
\begin{equation}
\frac{\beta}{H_*}
=
\left.
T\frac{d(S_3/T)}{dT}
\right|_{T_*}
\end{equation}
in a bounce-derived calculation.

In the present phenomenological scan, however, $\beta/H_*$ is treated as
an external input rather than being derived from the thermal bounce
action. Its influence on the gravitational-wave observables follows
directly from the adopted acoustic-source parameterization,
\begin{equation}
f_{\rm peak}
\propto
\frac{\beta}{H_*}
\frac{T_*}{v_w},
\end{equation}
and
\begin{equation}
\Omega_{\rm GW}^{\rm peak}h^2
\propto
\left(\frac{\beta}{H_*}\right)^{-1}
\left[
\frac{\kappa_{\rm sw}\alpha}
{1+\alpha}
\right]^2v_w.
\end{equation}

Thus, for fixed $T_*$, $v_w$ and transition strength, a faster transition
corresponds to a larger $\beta/H_*$ and tends to shift the characteristic
signal towards higher frequencies while reducing its peak amplitude.

The dependence of the peak frequency on the phenomenological inverse
transition duration is illustrated in Fig.~\ref{fig:multicolour_scan},
particularly in panel~(h), where the numerical points exhibit the expected
increasing trend. The same dependence is visible in the
frequency--amplitude plane shown in Fig.~\ref{fig:gw_detector_plane}.
Increasing $\beta/H_*$ shifts the characteristic frequency towards higher
frequencies, while, for otherwise comparable transition parameters, the
peak amplitude is reduced.

\begin{figure}[t]
    \centering
    \includegraphics[width=\textwidth]{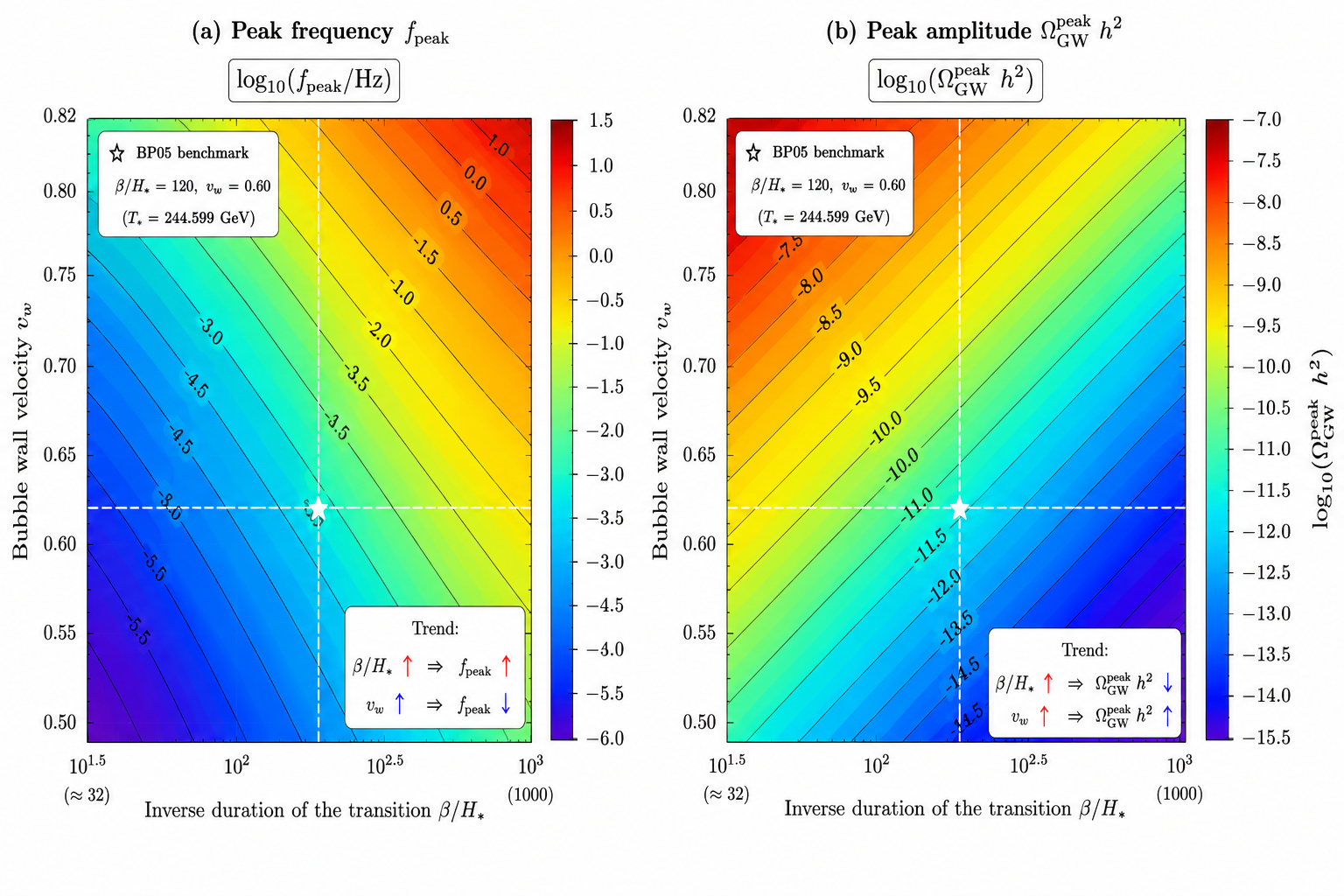}
    \caption{Sensitivity of the gravitational-wave observables to the
    inverse transition duration $\beta/H_*$ and the bubble-wall velocity
    $v_w$. The left panel shows the peak frequency
    $\log_{10}(f_{\rm peak}/{\rm Hz})$, while the right panel shows the
    peak gravitational-wave amplitude
    $\log_{10}(\Omega_{\rm GW}^{\rm peak}h^2)$ for the representative
    benchmark point BP05. The remaining parameters are fixed to their
    BP05 values, with $T_*=244.599$ GeV and $\alpha=0.03175$. The white
    dashed lines indicate the benchmark values
    $\beta/H_*=120$ and $v_w=0.60$. Increasing $\beta/H_*$ shifts the
    characteristic frequency towards higher frequencies while reducing
    the peak amplitude, whereas increasing $v_w$ lowers the peak
    frequency and enhances the peak amplitude. The figure therefore
    illustrates the explicit sensitivity of the gravitational-wave
    observables to the phenomenological transition-duration and
    bubble-wall-velocity parameters.}
    \label{fig:beta_vw_sensitivity}
\end{figure}

To further examine the combined influence of the phenomenological
parameters governing the transition duration and bubble expansion, we
consider the joint dependence on $\beta/H_*$ and $v_w$.
Figure~\ref{fig:beta_vw_sensitivity} presents this dependence for the
representative benchmark point BP05.

The left panel shows that $f_{\rm peak}$ increases with
$\beta/H_*$ and decreases with $v_w$, consistent with
\begin{equation}
f_{\rm peak}
\propto
\frac{\beta}{H_*}
\frac{T_*}{v_w}.
\end{equation}

The right panel shows the corresponding behaviour of the peak amplitude.
For fixed transition strength, increasing $\beta/H_*$ suppresses the
signal through the factor $H_*/\beta$, whereas increasing $v_w$ enhances
the acoustic contribution. These trends provide a direct visualization
of the parameter dependence adopted in the phenomenological scan.

\begin{figure}[t]
    \centering
    \includegraphics[width=\textwidth]{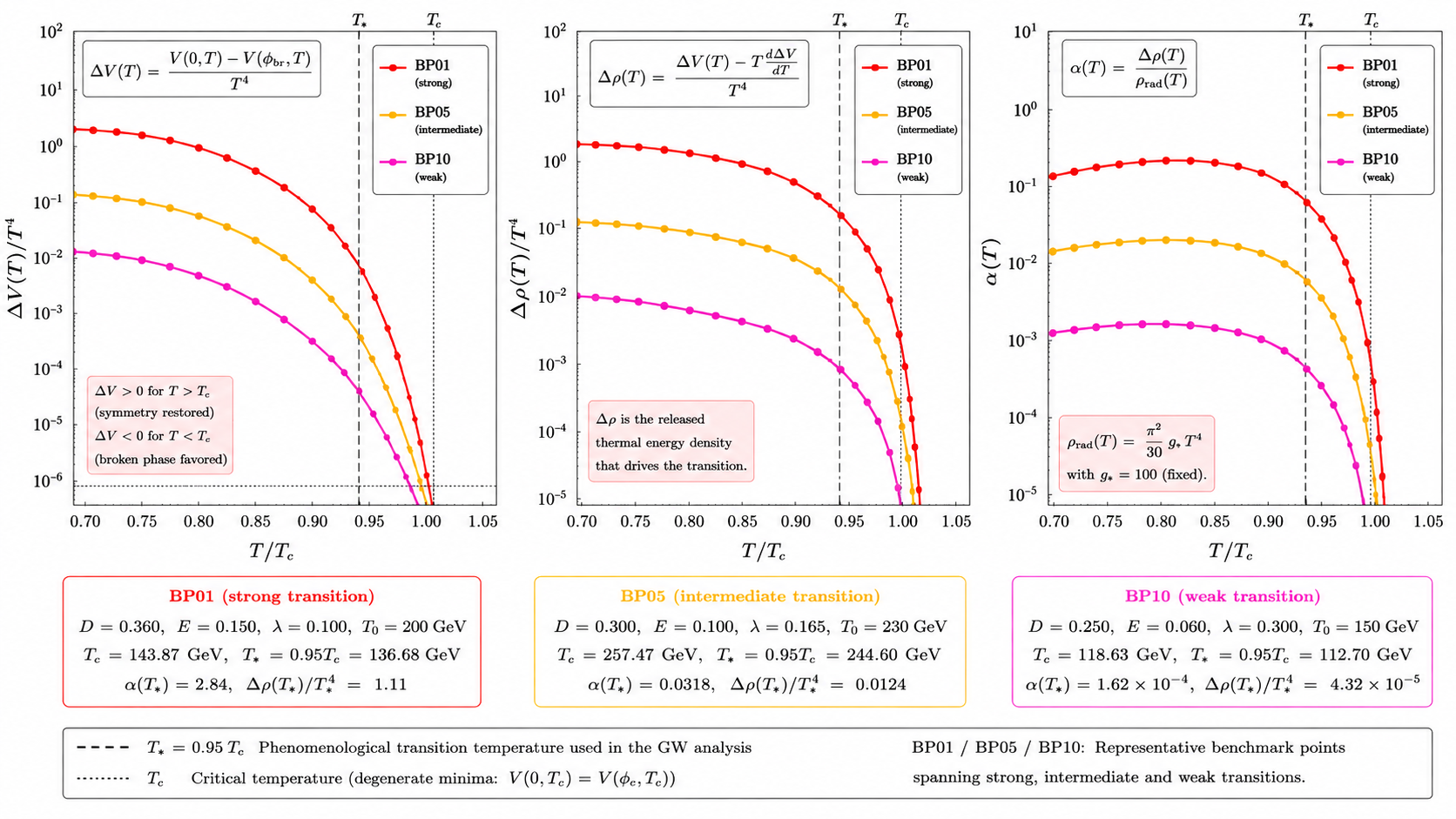}
    \caption{Thermodynamic evolution across the phase transition for the representative benchmark points BP01, BP05 and BP10. The three panels show, respectively, the dimensionless vacuum-energy difference $\Delta V(T)/T^4$, the dimensionless released energy density $\Delta\rho(T)/T^4$, and the transition-strength parameter $\alpha(T)=\Delta\rho(T)/\rho_{\rm rad}(T)$ as functions of $T/T_c$. The vertical dashed and dotted lines indicate the phenomenological transition temperature $T_*=0.95T_c$ and the critical temperature $T_c$, respectively. The figure illustrates the temperature dependence of the thermodynamic quantities entering the gravitational-wave calculation and the increasing transition strength across the representative benchmark points, with BP10 exhibiting the largest value of $\alpha$.}
    \label{fig:thermodynamic_evolution}
\end{figure}

The temperature dependence of the thermodynamic quantities entering the
gravitational-wave calculation is illustrated in
Fig.~\ref{fig:thermodynamic_evolution}. We show the evolution of the
vacuum-energy difference, released energy density, and transition
strength for three representative benchmark points spanning different
transition strengths.

For the stronger transition represented by BP10, the energy-density
difference and the corresponding transition strength are larger than for
BP01 and BP05. At the phenomenological transition temperature
$T_*=0.95T_c$, the three benchmark points therefore span a range of
$\alpha$, illustrating the sensitivity of the gravitational-wave amplitude
to the thermodynamic strength of the transition. The critical temperature
corresponds to the degeneracy condition
\begin{equation}
V(0,T_c)=V(\phi_c,T_c),
\end{equation}
while the use of $T_*=0.95T_c$ follows the phenomenological prescription
adopted in the present analysis.

\begin{figure}[t]
    \centering
    \includegraphics[width=\textwidth]{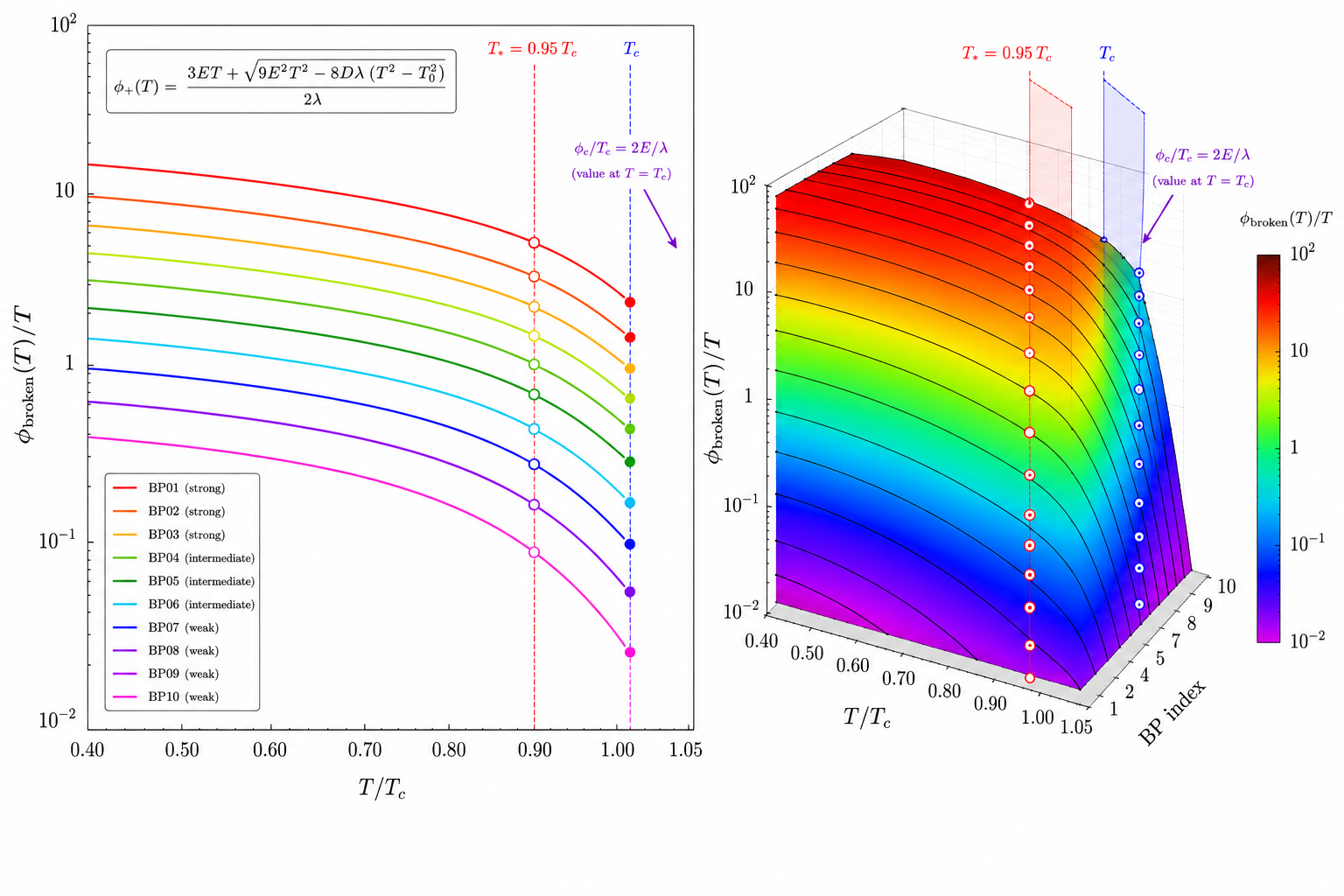}
    \caption{Temperature evolution of the broken-phase field value for
    the benchmark points BP01--BP10. The left panel shows
    $\phi_+(T)/T$ as a function of $T/T_c$, where $\phi_+(T)$ is the
    nonzero stationary solution of the finite-temperature effective
    potential. The right panel provides a two-dimensional representation
    of the same evolution across the benchmark points. The vertical
    reference lines indicate the critical temperature $T_c$ and the
    phenomenological transition temperature $T_*=0.95T_c$. At the
    critical temperature, the broken-phase field value satisfies
    $\phi_c/T_c=2E/\lambda$. The figure illustrates the variation of
    the broken-phase trajectory among the benchmark parameter points
    and its dependence on the underlying parameters $D$, $E$, $\lambda$,
    and $T_0$.}
    \label{fig:broken_phase_trajectory}
\end{figure}

The temperature evolution of the broken-phase field value provides
another useful characterization of the transition. Using the analytic
nonzero stationary solution,
\begin{equation}
\phi_+(T)=
\frac{
3ET+\sqrt{9E^2T^2-8D\lambda(T^2-T_0^2)}
}{2\lambda},
\end{equation}
we examine the dimensionless quantity $\phi_+(T)/T$ as a function of
$T/T_c$ for the ten benchmark points BP01--BP10. The resulting
trajectories are shown in Fig.~\ref{fig:broken_phase_trajectory}.
The benchmark trajectories exhibit a clear separation in the
dimensionless broken-phase field value. The benchmark points with larger
transition strength $\alpha$ generally maintain larger values of
$\phi_+(T)/T$ near the critical region, whereas the weaker transitions
approach smaller values as $T\rightarrow T_c$. This behaviour follows
directly from the temperature-dependent stationary solution and provides
a complementary view of the phase structure underlying the
gravitational-wave phenomenology.

The peak quantities discussed above characterize the location and
overall strength of the predicted gravitational-wave signal, but do not
display its spectral shape. To provide a more complete representation
of the predicted signal, we reconstruct the acoustic gravitational-wave
spectrum for representative benchmark points. Following the
sound-wave contribution, we write
\begin{equation}
\Omega_{\rm GW}(f)h^2
=
\Omega_{\rm GW}^{\rm peak}h^2
S_{\rm sw}(f),
\label{eq:gw_spectrum}
\end{equation}
where $S_{\rm sw}(f)$ denotes the normalized spectral shape of the
acoustic gravitational-wave contribution,
\begin{equation}
S_{\rm sw}(f)
=
\left(\frac{f}{f_{\rm peak}}\right)^3
\left[
\frac{7}
{4+3\left(f/f_{\rm peak}\right)^2}
\right]^{7/2}.
\label{eq:sw_shape}
\end{equation}
Thus, the peak frequency and peak amplitude are those obtained from
the phenomenological expressions discussed above, while
$S_{\rm sw}(f)$ determines the frequency dependence away from the
peak.

\begin{figure}[t]
    \centering
    \includegraphics[width=\textwidth]{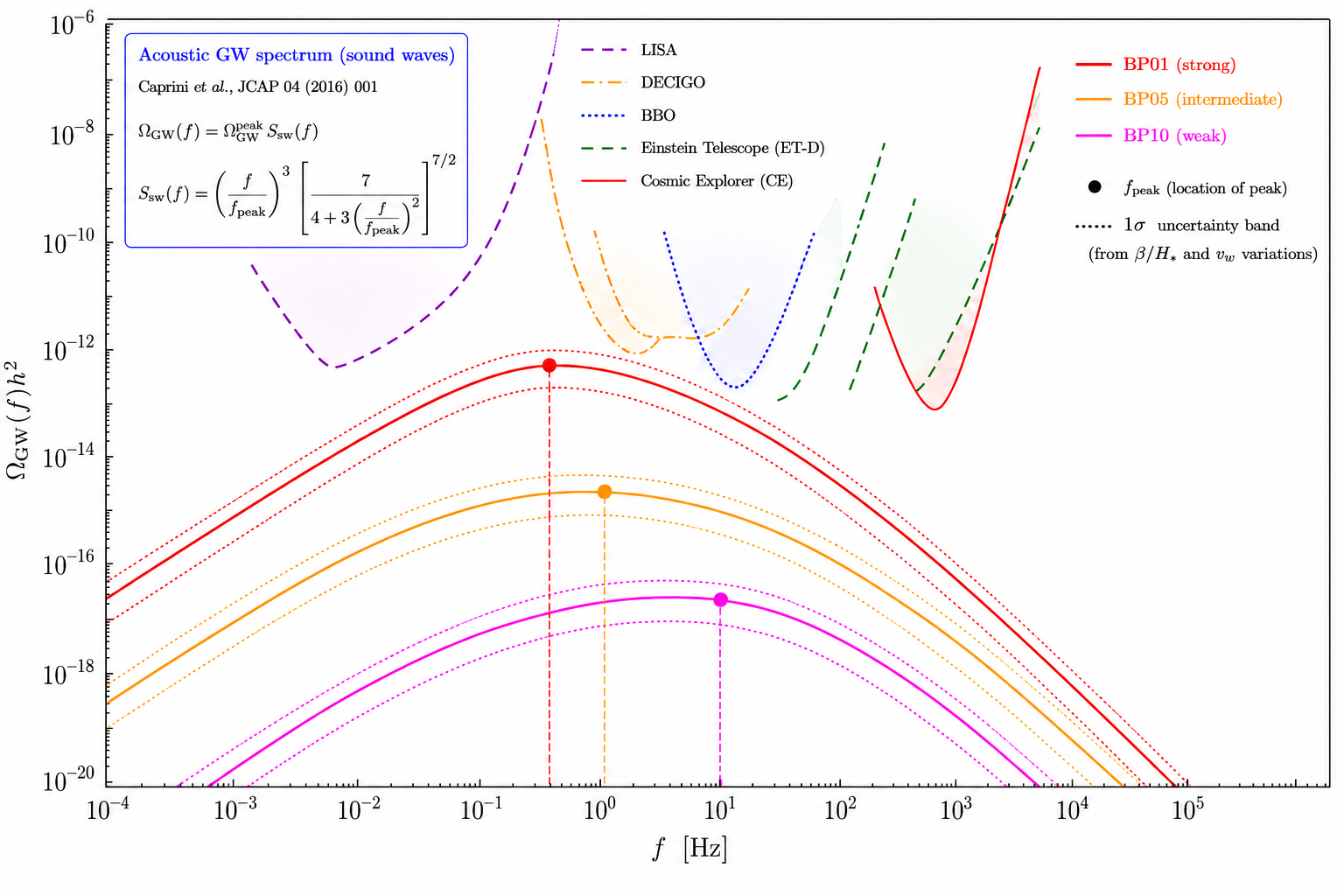}
    \caption{Acoustic gravitational-wave spectra for the representative
    benchmark points BP01, BP05 and BP10. The spectra are shown as
    $\Omega_{\rm GW}(f)h^2$ as a function of frequency $f$. The peak
    positions are determined by the phenomenological peak-frequency
    relation, while the normalization is fixed by
    $\Omega_{\rm GW}^{\rm peak}h^2$. The frequency dependence is
    described by the sound-wave spectral shape
    $S_{\rm sw}(f)$ given in Eq.~\eqref{eq:sw_shape}. The vertical
    markers indicate the corresponding peak frequencies. The
    representative detector sensitivity curves for LISA, DECIGO, BBO,
    ET-D and CE are included for phenomenological comparison with the
    predicted spectra.}
    \label{fig:gw_spectra}
\end{figure}
The three representative spectra illustrate the dependence of the
gravitational-wave signal on the transition strength. We consider the
benchmark points BP01, BP05 and BP10 defined in
Table~\ref{tab:benchmarks}. These points span increasing values of the
transition-strength parameter, with
$\alpha_*=0.02642$, $0.03175$, and $0.03940$, respectively.
Thus, BP01 represents the weakest transition among the three selected
benchmarks, while BP10 represents the strongest.

For the purpose of constructing the illustrative spectra, we take the
phenomenological parameters
\begin{equation}
\frac{\beta}{H_*}=100,
\qquad
v_w=0.60,
\end{equation}
for all three benchmark points. The corresponding transition
temperatures are those listed in Table~\ref{tab:benchmarks}, namely
$T_*=96.551$ GeV, $244.599$ GeV, and $1110.904$ GeV for BP01, BP05,
and BP10, respectively.

The peak locations are controlled by the inverse transition duration,
transition temperature, and bubble-wall velocity through
\begin{equation}
f_{\rm peak}\propto
\frac{\beta}{H_*}\frac{T_*}{v_w},
\end{equation}
while the peak amplitude depends nonlinearly on the transition strength
through
\begin{equation}
\Omega_{\rm GW}^{\rm peak}h^2
\propto
\left(\frac{H_*}{\beta}\right)
\left[
\frac{\kappa_{\rm sw}\alpha_*}{1+\alpha_*}
\right]^2
v_w.
\end{equation}

For the common choices $\beta/H_*=100$, $v_w=0.60$, and
$g_*=106.75$, the corresponding peak frequencies are
\begin{equation}
f_{\rm peak}
=
3.09\times10^{-3}\ {\rm Hz},
\quad
7.83\times10^{-3}\ {\rm Hz},
\quad
3.56\times10^{-2}\ {\rm Hz},
\end{equation}
for BP01, BP05, and BP10, respectively.

The corresponding peak amplitudes obtained from the phenomenological
acoustic-source expression are
\begin{equation}
\Omega_{\rm GW}^{\rm peak}h^2
=
1.21\times10^{-14},
\quad
2.46\times10^{-14},
\quad
5.62\times10^{-14},
\end{equation}
for BP01, BP05, and BP10, respectively.

\begin{table}[t]
\centering
\caption{Representative benchmark points used for the gravitational-wave
spectra shown in Fig.~\ref{fig:gw_spectra}. The values of
$\beta/H_*$ and $v_w$ are phenomenological inputs chosen as
$\beta/H_*=100$ and $v_w=0.60$ for the illustrative spectra.}
\label{tab:gw_spectral_benchmarks}
\begin{tabular}{c c c c c}
\hline
BP & $T_*$ [GeV] & $\alpha_*$ & $\beta/H_*$ & $v_w$ \\
\hline
BP01 & 96.551  & 0.02642 & 100 & 0.60 \\
BP05 & 244.599 & 0.03175 & 100 & 0.60 \\
BP10 & 1110.904 & 0.03940 & 100 & 0.60 \\
\hline
\end{tabular}
\end{table}

\begin{table}[t]
\centering
\caption{Peak gravitational-wave observables for the representative
benchmark points used in Fig.~\ref{fig:gw_spectra}. The peak
frequencies are calculated using Eq.~\eqref{eq:fpeak} with
$g_*=106.75$, $\beta/H_*=100$, and $v_w=0.60$.}
\label{tab:gw_peaks}
\begin{tabular}{c c c}
\hline
BP & $f_{\rm peak}$ [Hz]
& $\Omega_{\rm GW}^{\rm peak}h^2$ \\
\hline
BP01 & $3.09\times10^{-3}$ & $1.21\times10^{-14}$ \\
BP05 & $7.83\times10^{-3}$ & $2.46\times10^{-14}$ \\
BP10 & $3.56\times10^{-2}$ & $5.62\times10^{-14}$ \\
\hline
\end{tabular}
\end{table}
The comparison demonstrates that the representative benchmark points
occupy distinct regions of the frequency--amplitude plane. Since
$\beta/H_*$ and $v_w$ are held fixed for these three illustrative
spectra, the different peak frequencies arise primarily from the
different transition temperatures, through
$f_{\rm peak}\propto T_*$. The increase in transition strength from
BP01 to BP10 leads to a corresponding increase in the peak amplitude
through the efficiency factor $\kappa_{\rm sw}$ and the nonlinear
dependence on $\alpha_*$. The detector sensitivity curves shown in
Fig.~\ref{fig:gw_spectra} are included only to provide phenomenological
context for the frequency and amplitude ranges explored by the model.
They should not be interpreted as a precision detectability or
signal-to-noise calculation, since $\beta/H_*$ and $v_w$ are treated as
phenomenological inputs in the present analysis.
\section{Toward a first-principles phase-transition calculation}
\label{sec:limitations}

The present work is intentionally constructed as a phenomenological
numerical baseline. A complete first-principles prediction would require
replacing the phenomenological transition-temperature prescription by a
calculation of the thermal nucleation rate.

The tunnelling rate per unit volume can be written approximately as
\begin{equation}
\Gamma(T)
\simeq
T^4
\left(
\frac{S_3(T)}{2\pi T}
\right)^{3/2}
e^{-S_3(T)/T},
\label{eq:gamma}
\end{equation}
where $S_3(T)$ is the three-dimensional Euclidean bounce action.

The corresponding $O(3)$-symmetric bounce profile satisfies
\begin{equation}
\frac{d^2\phi}{dr^2}
+
\frac{2}{r}\frac{d\phi}{dr}
=
\frac{\partial V}{\partial\phi},
\label{eq:bounceeq}
\end{equation}
with boundary conditions
\begin{equation}
\left.
\frac{d\phi}{dr}
\right|_{r=0}
=
0,
\qquad
\phi(r\rightarrow\infty)
=
\phi_{\rm false}.
\end{equation}

The nucleation temperature can then be determined from the integrated
nucleation probability, rather than from the fixed phenomenological
prescription $T_*=0.95T_c$. The inverse transition duration can
subsequently be obtained from
\begin{equation}
\frac{\beta}{H_*}
=
T_*
\frac{d(S_3/T)}{dT}
\bigg|_{T_*}.
\label{eq:beta_bounce}
\end{equation}

Modern numerical treatments such as PhaseTracer2 provide tools for
connecting finite-temperature effective potentials with phase-transition
and gravitational-wave observables
\cite{Athron2025}.

A second improvement concerns the bubble-wall velocity. In the present
analysis, $v_w$ is treated as a phenomenological scan parameter. A
microscopic calculation would instead determine the wall velocity from
the balance between the driving pressure and the friction exerted by the
thermal plasma.

Finally, the acoustic contribution should be treated with care for
strong transitions. The finite lifetime of the acoustic source and
nonlinear plasma evolution can modify the gravitational-wave amplitude
relative to the simplest stationary-source estimate
\cite{RoperPol2024,Caprini2025}.

These extensions would provide a more complete first-principles treatment
of the nucleation dynamics, bubble-wall propagation, and gravitational-wave
production. The present results should therefore be regarded as a
phenomenological parameter-space baseline that can be systematically
refined by incorporating these effects.
\section{Conclusions}
\label{sec:conclusion}

We have investigated the thermodynamics and gravitational-wave phenomenology
of a first-order cosmological phase transition described by the
finite-temperature potential
\begin{equation}
V(\phi,T)
=
D(T^2-T_0^2)\phi^2
-
ET\phi^3
+
\frac{\lambda}{4}\phi^4.
\end{equation}
For each parameter point, we determined the critical temperature, the
broken-phase minimum, the phenomenological transition temperature
$T_*=0.95T_c$, the released energy density and the transition-strength
parameter $\alpha$, and subsequently calculated the corresponding
gravitational-wave peak frequency and amplitude. A dense numerical scan of
30,000 valid parameter points was performed to resolve the correlations
between the phase-transition thermodynamics and the gravitational-wave
observables.

A central result of the analysis is the identification of the dimensionless
combination
\begin{equation}
\xi=\frac{E^2}{D\lambda},
\end{equation}
which provides an analytic control parameter for the critical-temperature
shift,
\begin{equation}
\frac{T_c}{T_0}=\frac{1}{\sqrt{1-\xi}}.
\end{equation}
The numerical scan demonstrates how this analytic dependence propagates
through the thermodynamic quantities and into the gravitational-wave
frequency--amplitude plane. While $\xi$ organizes the critical-temperature
dependence, the transition strength and gravitational-wave observables retain
sensitivity to the remaining independent potential parameters as well as to
the phenomenological inputs $\beta/H_*$ and $v_w$. The representative
benchmark points BP01, BP05 and BP10 illustrate the corresponding variation
in the thermodynamic evolution and gravitational-wave properties.

The present calculation is deliberately phenomenological: $\beta/H_*$ and
$v_w$ are treated as independent inputs rather than being obtained from a
thermal bounce calculation or a microscopic treatment of bubble-wall
dynamics. Likewise, the prescription $T_*=0.95T_c$ is adopted instead of a
dynamical nucleation or percolation calculation. Within these assumptions,
the 30,000-point atlas provides a systematic mapping of the model parameter
space and identifies regions with potentially interesting gravitational-wave
signals. A natural extension is to determine $S_3(T)/T$, the nucleation and
percolation temperatures, $\beta/H_*$ and $v_w$ dynamically, together with
an improved treatment of the finite lifetime and nonlinear evolution of the
acoustic source. This would allow the analytic--numerical framework developed
here to be extended toward a fully dynamical description of the phase
transition.

\section*{Data availability}

The numerical dataset and the code used to generate the results and
figures will be made publicly available in a repository upon publication.
\bibliographystyle{JHEP}

\end{document}